\documentclass[twocolumn, secnumarabic,amssymb, nobibnotes, aps, pre,groupedaddress,10pt]{revtex4-1}
\usepackage{color}

\usepackage[utf8x]{inputenc}
\usepackage{enumerate}
\usepackage{enumitem}
\usepackage{amsmath}
\usepackage{amssymb}
\usepackage{graphicx}
\usepackage{appendix}
\usepackage{amsthm}
\usepackage{subfigure}

\newcommand{\dy}{{\Delta}y}
\newcommand{\dt}{{\Delta}t}

\newcommand{\ensav}[2]{\left\langle #1 \right\rangle_{#2}}

\newcommand{\dsum}{\displaystyle\sum}

\newcommand{\mybar}[1]{ ~\overline{#1} }

\numberwithin{equation}{section}

\begin{document}

\title{The Effect of Nonlinearity in Hybrid KMC-Continuum models.}

% \author{Ariel Balter}
%    \affiliation{Pacific Northwest National Laboratory P.O. Box 999, Richland, WA 99352}
%    \email{ariel.balter@pnl.gov}
% \author{Guang Lin}
%    \affiliation{Pacific Northwest National Laboratory P.O. Box 999, Richland, WA 99352}
%    \email{alexandre.tartakovsky@pnl.gov}
% \author{Aleaxndre M. Tartakovsky}
%    \affiliation{Pacific Northwest National Laboratory P.O. Box 999, Richland, WA 99352}
%    \email{alexandre.tartakovsky@pnl.gov}

\author{Ariel Balter, Guang Lin, and Alexandre M. Tartakovsky}

\begin{abstract}
Recently there has been interest in developing efficient ways to model heterogeneous surface reactions with hybrid computational models that couple a KMC model for a surface to a finite difference model for bulk diffusion in a continuous domain.  We consider two representative problems that validate a hybrid method and also show that this method captures the combined effects of nonlinearity and stochasticity.  We first validate a simple deposition/dissolution model with a linear rate showing that the KMC-continuum hybrid agrees with both a fully deterministic model and its analytical solution.  We then study a deposition/dissolution model including competitive adsorption, which leads to a nonlinear rate, and show that, in this case, the KMC-continuum hybrid and fully deterministic simulations do not agree.  However, we are able to identify the difference as a natural result of the stochasticity coming from the KMC surface process.  Because KMC captures inherent fluctuations, we consider it to be more realistic than a purely deterministic model.  Therefore, we consider the KMC-continuum hybrid to be more representative of a real system.
\end{abstract}

\maketitle

\section{Introduction}
Kinetic Monte Carlo (KMC) is an established stochastic method for simulating dynamic, spatially inhomogeneous surface phenomena at the atomistic level.  This method samples the master equation, a probabilistic description of surface processes, such as reactions, deposition/dissolution, diffusion, etc. \cite{Chatterjee2007}.  KMC is advantageous for predicting or understanding experimentally observed surface processes, such as morphology \cite{meakin2008} or reaction rates \cite{Chatterjee2004}, that are complex and do not permit analytical solutions.  One important application is in modeling catalytic reactors \cite{Dudukovic2009,Neurock2004,Mei2010}.  The stochastic nature of the processes involved can drive emergent behavior. It is well known that in systems containing nonlinearities, microscopic fluctuations can instigate both microscopically and macroscopically observed behavior that would not be predicted in a deterministic model \cite{Nicolis1971,vankampen2007spp,Ortoleva1987,0305-4470-19-3-008}.  One can construct mean field models (deterministic models for the average behavior of the surface), but these cannot capture the full complexity of a surface process.  For nonlinear systems, a purely deterministic mean field model may not even predict the ensemble average behavior.

For the most part, KMC has been used to model processes on (or in) a solid interface, and only recently have there been a few attempts to couple a KMC surface model to bulk phase diffusion, for instance, a surface immersed in a solution with a reactive species, or other multi-scale models.  Saedi, Drews et al., and Pricer et al. modeled electrodeposition of copper \cite{Drews2005,pricer2002,Saedi2006,Saedi2006b} and Mei and Lin modeled the CO oxidation over a Ruthenium catalyst substrate \cite{Mei201156}.  KMC models of surface kinetics are known show different behavior than mean field models of the same surface.  Therefore, it is very reasonable that a coupled KMC/diffusion model could show different behavior than a mean field model for the same system.  In the previous works, the purpose of the KMC portion of the model is to generate realistic surface kinetics that supply the bulk diffusion portion of the model with a realistic surface energy \cite{Saedi2006,Saedi2006b} or turnover rate \cite{Mei201156}.  Saedi did compare a coupled KMC/diffusion model to a mean field model with a deterministic surface process and demonstrated that these to models produce different results.  However, he did not address the origin of this difference.  Our goal is to focus on the effect of fluctuations on nonlinear rate laws on a more fundamental level.

Real surface processes are very complex.  Adsorption, desorption, surface diffusion, and reactions all alter spatial distribution and even the contour of a surface. By altering the immediate neighborhood of each site these processes, in turn, alter the free energy of each site which affects future surface evolution.  Furthermore, each of these processes evolves randomly.  A mean field model for the surface, especially a deterministic one, can not include this complexity.  Therefore, one would expect that a model that couples diffusion to a deterministic, mean-field model of surface evolution would have very different behavior than a \emph{hybrid} that couples diffusion to a more realistic KMC model.  However, it is not immediately apparent if we can understand at a theoretical level what the differences would be and how they would arise.  Towards this end, we study two representative problems, one with a linear rate law and one with a nonlinear rate law.  In both problems, an adsorption/desorption process exchanges a single species between a solid phase ($s$) localized to a surface and a gas-phase ($g$) that diffuses freely in a continuous domain:

\begin{equation}
\label{eqn:surfrx}
s \rightleftarrows g.
\end{equation}

Deposition and dissolution occur when a solid phase is in contact with a solution.  Individual molecules may deposit onto the solid or dissolve into the solution.  When this happens, the surface acquires velocity, growing or receding in height.  This velocity may be spatially heterogeneous, leading to spatially heterogeneous surface morphology. In our simple model, we ignore the shape and velocity of the surface, making the surface an infinite source and sink -- there is always plenty of room for deposition and an unlimited supply for dissolution.  The rate equation for this model is linear.

In catalysis, one or more reactant species adsorb to a fixed surface of a different species, the catalyst.  The catalyst facilitates the transformation of these species into one or more product species, which may desorb back into the solution.  However, the catylist itself remains intact.  Because a finite surface has only a finite number of binding sites (in our model, $s_0$), the surface can ``fill up'', leading to competitive adsorption \cite{Masel1996}.  Under these conditions, the rate law has a nonlinear term.

In both of our models we solve the diffusion equation with the same deterministic forward Euler finite difference (FEFD) algorithm.  However, we evolve the surface process in two different ways: (1) a deterministic FEFD, and (2) a stochastic master equation solved using the kinetic Monte Carlo (KMC) algorithm.  The surface process model (KMC or mean field) supplies the boundary condition at the surface.

We validate our numerical methods using the linear  model (deposition/dissolution) which has an analytical solution for 1D diffusion with a deterministic boundary condition corresponding to the mean field model.  We find perfect agreement between the analytic, mean field numerical and KMC hybrid models. We then compare the numerical solutions of both the mean field model, and the KMC hybrid model, for the system with a nonlinear surface rate.  For a wide range of parameters, we find that the mean field model overestimates the average surface occupancy, and this result agrees with our mathematical analysis.  In Sec. \ref{sec:math_model} we describe the mathematical model, a partial differential equation (PDE), that corresponds to the physical system we consider in this paper.  In Sec. \ref{sec:stoch_splitting} we present a moment equation analysis that shows how nonlinearity in the catalysis model alters the steady-state surface equation when stochasticity is present.  In Sec. \ref{sec:algorithms} we describe the numerical algorithms we use integrate the PDE described in Sec. \ref{sec:math_model}.  Sec. \ref{sec:results} contains the graphical results of our simulations.  We analyze and discuss these in Secs. \ref{sec:results} and \ref{sec:conclusion}.

\section{Mathematical Models \label{sec:math_model}}
Our model system consists of a 1D continuum domain with a reactive boundary condition at one end and a fixed concentration at the other.  In the continuum domain, gas particles diffuse leading to local changes in concentration.  At the reactive boundary, particles adsorb and desorb according to a rate law $R(s(t),g(0,t))$, where $s(t)$ is the number of particles \emph{on} the surface and $g(0,t)$ is the concentration \emph{at} the surface.  In the deposition/dissolution model, $s(t)$ and $g(y,t)$ are completely decoupled, $R(s(t),g(0,t)) \equiv R(g(0,t))$, so we do not need to explicitly track the surface concentration.  For the catalysis model, $s(t)$ and $g(y,t)$ are coupled.  KMC is a particle based algorithm (for evolving a particle based model), while the PDE models concentration.  So, except as otherwise noted, we refer to concentration in the continuum domain and particle number on the surface. 

\begin{figure*}
%  \centering
 \subfigure[System Modeled: $s_0$ only applies to catalysis model.]{
   \includegraphics[height=3in]{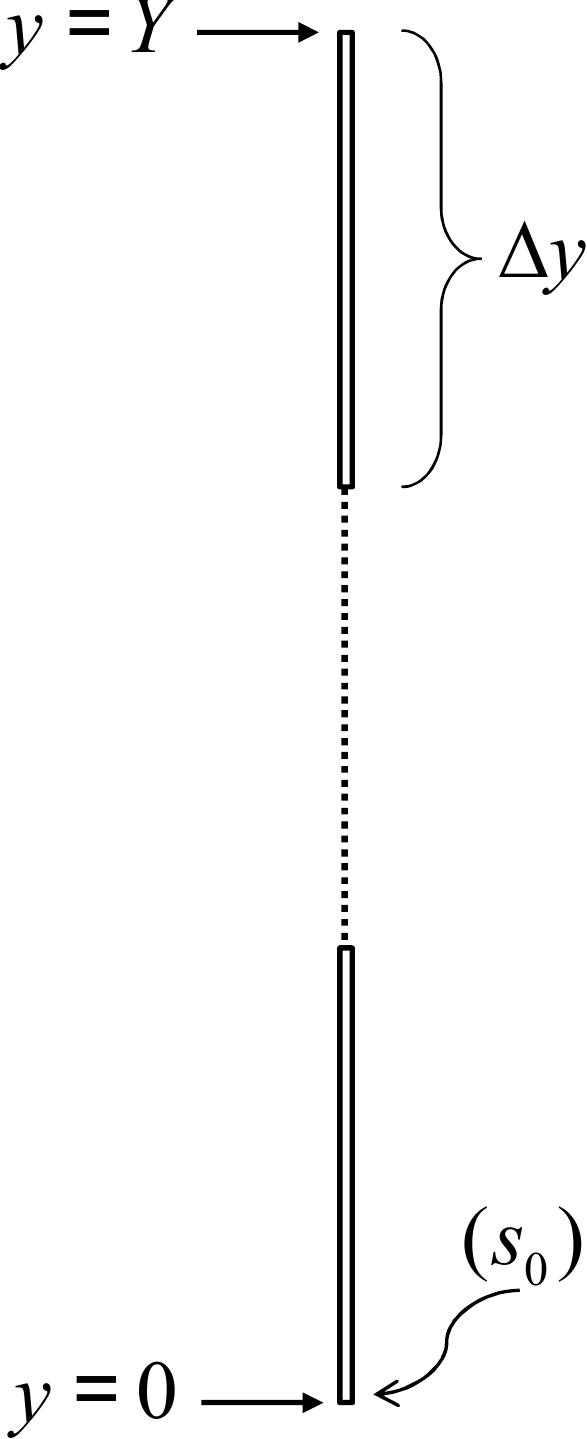}
   \label{subfig:modeled}
   }
\hspace{0.5in}
 \subfigure[Conceptualized System: $s_0$ only applies to catalysis model.]{
   \includegraphics[height=3in]{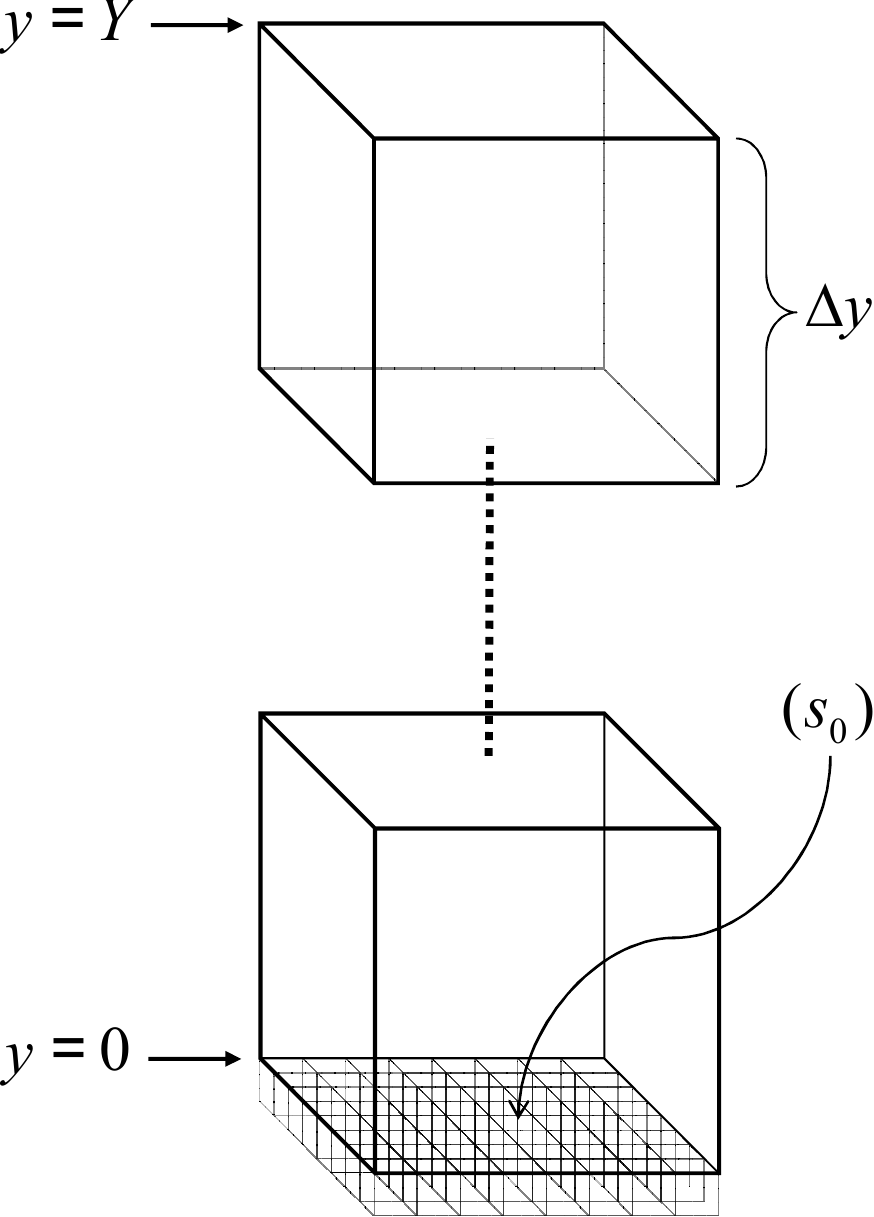}
   \label{subfig:conceptualized}
   }
 \caption{(a): Actual model -- 0D surface and 1D continuum domain.  (b): Conceptual model -- 2D surface and 3D continuum domain. $s_0$ only applies to catalysis model.\label{fig:domains}}
\end{figure*}

The system we model is a quite simplified version of a more realistic system consisting of a 2D surface bounding a 3D continuum domain. Only adsorption and desorption occur at the surface, and there are no neighbor interactions of any kind. As a result, the entire surface is reduced to a single 0D lattice site.  By having each gas phase voxel (grid cell) span the entire surface, we reduce the 3D gas-phase domain into a 1D domain discretized into a line of voxels.  We illustrate the conceptual model just described and the actual model in Fig. \ref{fig:domains}.  The finite number of lattice sites $s_0$ plays a role in our catalysis model, but not in the deposition/dissolution model.  However, the individual binding sites in Fig. \ref{subfig:conceptualized} are for conceptual clarity, and purely illustrative.

We can consider two ways to formulate a PDE model for diffusion with a reactive boundary: (1) We could impose a Robin boundary condition at the surface, or (2) Include a delta function ($\delta(y)$) source at the boundary and a Neumann (no flux) boundary condition. From a practical point of view, it is often simpler, more accurate \cite{Ryan201156}, and quite common in numerical work \cite{AlRawahi2002471} to use the second approach:  

\begin{subequations}
   \label{eqn:With_Source}
 \begin{equation}
  \label{eqn:PDE_With_Source}
  \frac{\partial g}{\partial t} = D \frac{\partial^2 g}{\partial x^2} - R(s(t),g(y,t))\delta(y)
 \end{equation}
 \begin{equation}
  \label{eqn:Left_BC_Source}
   -D\frac{\partial g}{\partial x} = 0 \text{   at   } y=0
 \end{equation}
\begin{equation}
  \label{eqn:Right_BC_Source}
    g(Y,t) = g_Y
 \end{equation}
 \begin{equation}\label{eqn:Surf_Rate_Source}
  \frac{ds}{dt} = R(s(t),g(0,t))
 \end{equation}
\end{subequations}

\noindent where $R(s(t),g(0,t)$ is the rate law describing the surface process.  Surface rate coefficients are specified in terms of inverse area \cite{Ryan201156}, while bulk rate coefficients are specified in terms of inverse volume.  The $\delta$ function has units of $V^{-1}$ and accounts for this.

\subsection{Linear Model \label{sec:linear_model}}
The linear model is representative of a surface growth process involving adsorption and desorption \cite{Tartakovsky2007a,tartakovsky2008,Tartakovsky2007}.  Such models may use a surface reaction of the form:

\begin{equation}
 \label{eqn:linear_source_rx}
R(s,g(0,t))  = k^{on}g(0,t) - k^{off} 
\end{equation}

\noindent The model, comprised of Eqs. \eqref{eqn:With_Source} and \eqref{eqn:linear_source_rx}, permits an analytic solution:

\begin{equation}
 \label{eqn:partial_full}
 g(y,t) = Py + Q + \dsum_{n=1}^N b_n \sin\left[\lambda_n(y-Y)\right]e^{-D\lambda_n^2 t}
\end{equation}

\noindent with

\begin{align}
 \label{eqn:auxiliary}
 \lambda_n = -\frac{k^{on}}{D} \tan \lambda_n Y \\ \nonumber
 P = \frac{k^{on}g_Y - k^{off}}{k^{on}Y+D} \\ \nonumber
 Q = \frac{k^{off}Y + Dg_Y}{k^{on}Y+D}
\end{align}

\noindent and

\begin{equation}
 \label{eqn:b_n}
b_n = \frac{4}{\lambda_n} \frac{ P \sin{\lambda_n Y} - P\lambda_n Y + Q \lambda_n \cos{\lambda_n Y} - Q \lambda_n}{\sin{2 \lambda_n Y} - 2 \lambda_n Y}
\end{equation}

We see that the steady-state, $t \rightarrow \infty$, solution is

\begin{equation} \label{eqn:linear_steady_state}
 g_{ss}(y) = \frac{k^{on}g_Y - k^{loff}}{k^{on}Y+D} y + \frac{k^{off}Y + Dg_Y}{k^{on}Y+D}
\end{equation}

Since fluctuations should not affect the ensemble mean in the linear model (see Sec. \ref{sec:stoch_splitting}), we use the linear model to validate our numerical methods.  When, for a range of parameters, we see a match between the analytic solution, the deterministic mean field model, and the ensemble average of many realizations of the hybrid KMC/diffusion model, then we assume this validates our numerical methods.

\subsection{Nonlinear Model}
In the nonlinear model,

\begin{align}
  \label{eqn:nonlin_source_rx}
  R(s,g(0,t)) & = \frac{ds}{dt} \\ \nonumber
 & = k^{on}g(0,t) \left( s_0 -s(t) \right) - k^{off}s(t)
\end{align}

\noindent We see that, in the nonlinear case, Eq. \eqref{eqn:With_Source} is coupled to the surface because, in this case, the surface rate is explicitly a function of the number of particles \emph{on} the surface $s(t)$ and the gas concentration \emph{at} the surface $g(0,t)$. When the number of particles on the surface  reaches a steady state, i.e. $ds/dt=0$, this constrains Eq. \eqref{eqn:With_Source} to:

\begin{subequations}
   \label{eqn:With_Source_SS}
 \begin{equation}
  \label{eqn:PDE_With_Source_SS}
  \frac{\partial g}{\partial t} = D \frac{\partial^2 g}{\partial x^2}
 \end{equation}
 \begin{equation}
  \label{eqn:Left_BC_Source_SS}
   -D\frac{\partial g}{\partial x} = 0 \text{   at   } y=0
 \end{equation}
\begin{equation}
  \label{eqn:Right_BC_Souce_SS}
    g(Y,t) = g_Y
 \end{equation}
\end{subequations}

\noindent The steady state $\partial g / \partial t=0$ solution of Eq. \eqref{eqn:With_Source_SS} gives $g(y,\infty) = g_Y$.  Therefore, using Eq. \eqref{eqn:nonlin_source_rx}, we can derive the steady state number of particles on the surface

\begin{equation}
 \label{eqn:s_eq_nonlin}
 s_{ss} = s_0\frac{k^{on} g_Y}{k^{on} g_Y + k^{off}}
\end{equation}

\section{Effect of fluctuations \label{sec:stoch_splitting}}
At the molecular scale, chemical processes are random processes.  Molecular scale models such as KMC include these random fluctuations.  Linear stochastic differential or partial-differential equations have the same ensemble mean as an equivalent deterministic equation.  However, nonlinear equations may not.  Using the method of moment equations \cite{Bolker1997}, we can sometimes predict the differences.

Suppose $x(t)$ is a stochastic function of time.  The method of moments begins with Reynolds decomposition where we divide $x(t)$ into a deterministic part and a stochastic part: Then $x(t) = \mybar x(t) + x'(t)$, where $\mybar x(t) = \ensav{x(t)}{\omega}$ is the ensemble average (over realizations $\omega$) of $x(t)$, and $x'(t)$ is the stochastic part.  Notice that our definition ensures $\mybar x'(t) = 0$. The Reynolds decomposition of Eq. \eqref{eqn:With_Source} is:

\begin{subequations}
   \label{eqn:With_Source_RD}
 \begin{multline} \label{eqn:PDE_With_Source_RD}
  \frac{\partial}{\partial t}(\mybar g(y,t) + g'(y,t)) = D \frac{\partial^2}{\partial y^2}(\mybar g(y,t) + g'(y,t)) \\
   - \left[ \mybar R\left( s(t),g(y,t) \right) + R'\left( s(t),g(y,t) \right) \right]\delta(y)
 \end{multline}
 \begin{equation} \label{eqn:Left_BC_Source_RD}
   -D\frac{\partial}{\partial y}(\mybar g(y,t) + g'(y,t)) =  0 \text{   at   } y=0 \\
 \end{equation}
\begin{equation} \label{eqn:Right_BC_Source_RD}
    \mybar{g}(Y,t) + g'(Y,t) = g_Y \\
 \end{equation}
 \begin{equation} \label{eqn:Surf_Rate_Source_RD}
  \frac{d}{dt}(\mybar{s}(t) + s'(t)) = \mybar R\left( s(t),g(0,t) \right) + R'\left( s(t),g(0,t) \right)  \\
 \end{equation}
\end{subequations}

\noindent On account of the fact that Eq. \eqref{eqn:With_Source} is entirely linear, the ensemble average returns exactly Eq. \eqref{eqn:With_Source}.

The Reynolds Decomposition of the linear surface rate Eq. \eqref{eqn:linear_source_rx} gives:

\begin{multline}\label{eqn:Linear_Source_RD}
   \mybar R\left( s(t),g(0,t) \right) + R'\left( s(t),g(0,t) \right) = \\
  k^{on}(\mybar{g}(0,t) + g'(0,t)) - k^{off}
\end{multline}

\noindent Again, the ensemble average of Eq. \eqref{eqn:Linear_Source_RD} is exactly Eq. \eqref{eqn:linear_source_rx}.  However, for the nonlinear surface rate, Eq. \eqref{eqn:nonlin_source_rx}, the Reynolds Decomposition gives:

\begin{multline}\label{eqn:Nonlin_Source_RD}
   \mybar R\left( s(t),g(0,t) \right) + R'\left( s(t),g(0,t) \right) = \\
k^{on}(\mybar{g}(0,t) + g'(0,t))(s_0 - \mybar{s}(t) - s'(t)) \\ 
  - k^{off}(\mybar{s}(t) + s'(t)) 
\end{multline}

\noindent Now the ensemble average includes a non-vanishing cross moment:

\begin{multline} \label{eqn:nonlin_pde_fluct}
 \mybar R\left( s(t),g(0,t) \right) = \\
   k^{on}\mybar{g}(0,t)(s_0 - \mybar{s}(t)) - k^{off}\mybar{s}(t) \\ 
 - k^{on}\mybar{g'(0,t)s'(t)}
\end{multline}

\noindent At steady state, $\mybar {g}_{ss}(0,\infty) = g_Y$ and $\frac{d}{dt}\mybar{s} = 0$, so:

\begin{equation}
 \label{eqn:ens_av_steady_state}
k^{on}(\mybar{s'g'})_{ss} = k^{on} g_Y (s_0 - \mybar s_{ss}) - k^{off} \mybar s_{ss}
\end{equation}

\begin{figure*}
 \centering
 \includegraphics[width=0.8\textwidth]{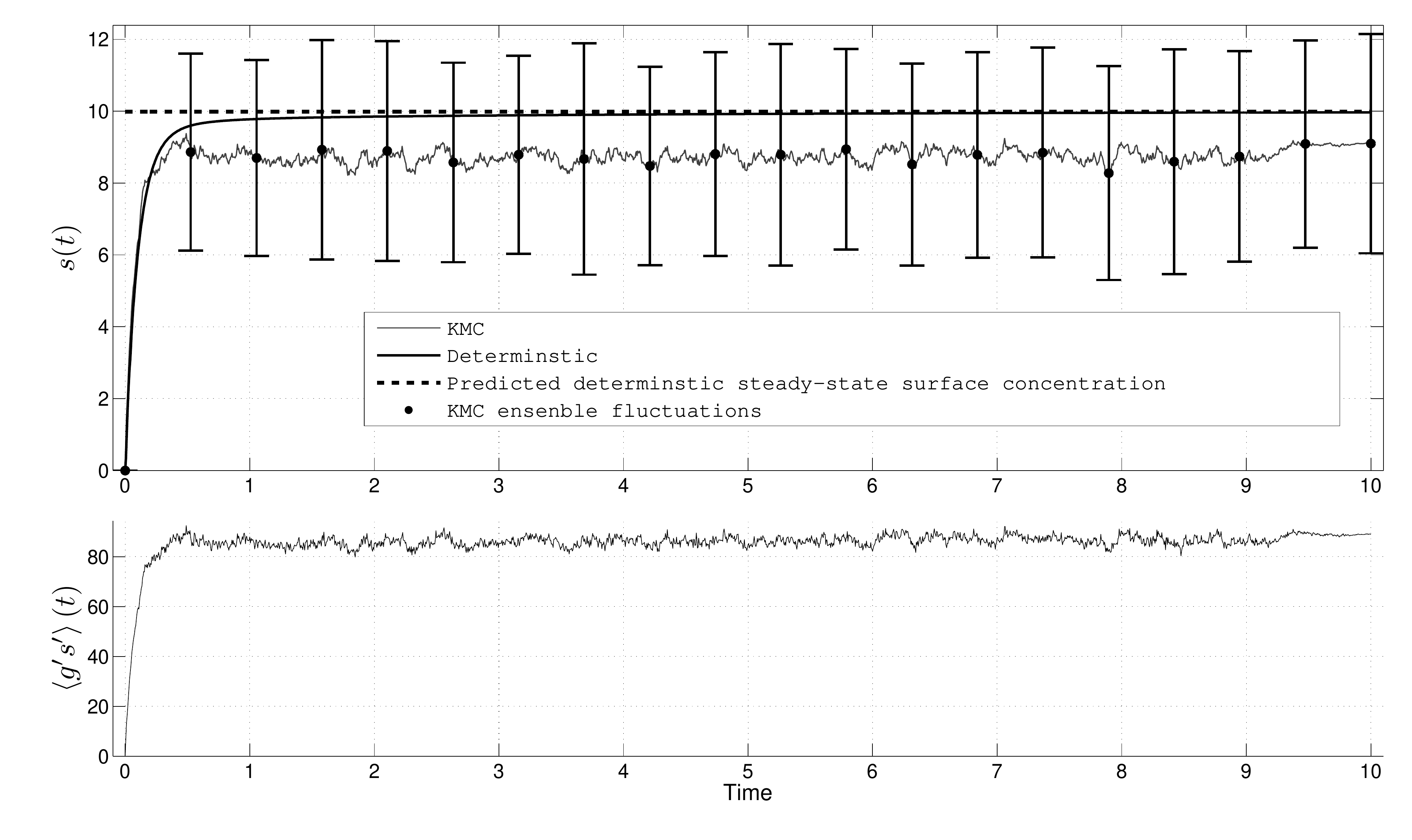}
 \caption{\label{fig:cov_neg_seq_less} In the lower panel we see that the cross moment $\mybar{s_{ss}'g_{ss}'}$ stabilizes to a value that is $>0$, and $s_{ss}^{stoch} < s_{ss}^{det}$, as predicted by Eq. \eqref{eqn:s_eq_stoch}. (Ensemble size: $N=200$, $nY=10^1$, $S_0=10^4$, $k^{on}=10^{-3}$, $k^{off}=10$, $Y=10^2$, $D=10^3$) }
\end{figure*}

\noindent Thus, when we consider fluctuations, we have a new steady state value for the number of particles on the surface:

\begin{equation}
 \label{eqn:s_eq_stoch}
 \mybar s_{ss} = \frac{k^{on} g_Y s_0 - k^{on}(\mybar{s'g'})_{ss}}{k^{on}g_Y + k^{off}}
\end{equation}

\noindent which can be either greater or less than the deterministic case depending on the sign of the cross moment $(\mybar{s'g'})_{ss}$.  It is challenging to derive analytical expressions for these cross moments.  However, we may approximate them numerically.  We can only approximate them due to the fact that fluctuations are rapid and individual realizations evolve in asynchronous time steps.  Thus it is impossible to calculate the ensemble statistics at an \emph{exact} point in time.  We estimate the ensemble statistics in the same way as we do for the ensemble statistics in our results, Sec. \ref{sec:results}.  In Fig. \ref{fig:cov_neg_seq_less} we have chosen parameters such that the cross moment $\mybar{s'g'}_{ss}$ is particularly large, leading to a clearly visible effect.  On the other hand, there are many parameter ranges where $\mybar{s'g'}_{ss}$ is small, and thus, so is the effect.  See Sec. \ref{sec:results} for examples.

\subsection{System Characterization}
% Both the linear and nonlinear system can be characterized by two dimensionless constants:
 \begin{subequations}\label{eqn:dimensionless_constants}
    \begin{equation}\label{eqn:q1}
 	   Q_1 = \frac{k^{on}s_0Y}{D}
  	\end{equation}
 	\begin{equation}\label{eqn:q2_1}
 		Q_2 = \frac{k^{off}s_0Y}{Dg^*}
 	\end{equation}
 \end{subequations}
The parameter $g^*$ is a characteristic gas-phase concentration.  These constants represent whether the on-reaction or off-reaction is reaction-limited or diffusion-limited.  In the mean-field model, a natural choice for $g^*$ is the fixed concentration $g_Y$.  However, in the master-equation model, we work with integral numbers of particles.  Since $g_Y$ is both the fixed boundary concentration and the equilibrium concentration, we will also use the representation:
 \begin{equation}\label{q2_2}
  Q_2 = \frac{k^{off}s_0Y^2}{Dn_Y}
 \end{equation}
\noindent where $n_Y=g_YL$ is the total number of particles in the domain when the gas concentration has reached steady-state. This is consistent with the units represented in our graphs.

% However, in the master-equation model, we work with integral numbers of particles.  Numerically (see section \ref{sec:algorithms}) we adjust the fixed boundary concentration, however, physically this represents an integral number of particles in the voxel at $y=Y$, i.e. $n_Y = g_Y\dy$.  To circumvent this dependence on our numerical discretization, we will 

\section{Algorithms \label{sec:algorithms}}

\subsection{Finite Difference continuum model for bulk diffusion \label{sec:finite_difference}}
To solve Eq. \eqref{eqn:With_Source} we used a standard FEFD algorithm.  However, in the KMC hybrid we let the KMC take the place of Eqs. \eqref{eqn:linear_source_rx} and \eqref{eqn:nonlin_source_rx}.  We discretize the space between $y=0$ and $y=Y$ into $N$ voxels centered at $\dy (i-1) $ where $i = 1...N$ and $\dy = Y/(N-1)$.  With this model, the numerical integration step for the gas-phase concentration is

\begin{subequations}
\begin{multline}
\label{eqn:bulk_diff_discretization}
g^{t+\dt}_i = g^{t}_i + D\frac{\dt}{\dy^2}\left[ g^{t}_{i+1} - 2g^{t}_i + g^{t}{i-1} \right] \\
 - R(s^t,g_0^t)\dt\delta_{i,0}/\dy \text{   for   } i = 1...N-1
\end{multline}
\begin{equation}
 \label{eqn:bulk_left_disc}
 -D\frac{g_1^t - g_0^t}{\dy} = 0  \implies  g_1^t = g_0^t
\end{equation}
\begin{equation}
 \label{eqn:bulk_right_disc_bc}
   g^{t}_N = g_Y
\end{equation}
\end{subequations}

\noindent We integrate the number of particles on the surface using

\begin{equation}
 \label{eqn:deterministic_R_fefd}
   s^{t+\dt} = s^t + R(s^t,g_0^t)\dt
\end{equation}

\noindent in which we have discretized the Dirac delta function $\delta(y)$ using the Kr\"{o}necker delta function $\delta_{i,0}$.

In the linear case (deposition/dissolution), the rate is

\begin{equation}\label{eqn:disc_linear_rx}
 R(s^t,g_0^t) = k^{on}g_0^t - k^{off}
\end{equation}

\noindent In the nonlinear case (catalysis), the rate is

\begin{equation}\label{eqn:disc_nonlinear_rx}
 R(s^t,g_0^t) = k^{on}g_0^t(s_0 - s^t) - k^{off}s^t
\end{equation}

\noindent For stability, we chose $\dt = \alpha \times \text{MIN}\{\dy^2/D, 1/(k^{on}g^t_{surf}), 1/k^{off})\}$, where $\alpha<1$ is a stability factor, usually $0.1$.  

\subsection{Kinetic Monte Carlo for the surface process \label{sec:kmc}}
Because one single bulk phase voxel spans the entire hypothetical surface, molecules in the voxel at $y=0$ have an equal likelihood of interacting with any surface molecule or empty site.  Likewise, because each surface molecules reacts only with gas molecules, and not other surface molecules, surface molecules each have the same probability of interacting with any gas molecule.  Therefore, we treat the ``surface'' as a zero-dimensional point located at $y=0$.  Traditionally, one would use kinetic Monte Carlo (KMC) to generate individual realizations of the master equation corresponding to spatially inhomogeneous surface process.  Since we have homogenized our surface, a small volume that encompasses the surface ($s^t$) and adjacent bulk phase $g_0^t$ constitute a "well-mixed" system.  In this respect, our KMC reduces to the simpler Gillespie algorithm for a small, well-mixed volume containing both surface and an adjacent small volume of the bulk phase.  We will continue using the term KMC so as to remind the reader that we are using it for a surface process.

We employ the "next reaction" form of this algorithm, which we will briefly review.  We have \emph{on} and \emph{off} transition rates that are re-defined at each time step according to:

\vspace{12pt}
\begin{tabular*}{0.45\textwidth}{ l  c r }
\hline \\
   & $\mathbf{r_{on}}$ & $\mathbf{r_{off}}$ \\  [12pt]
  \textbf{Linear\sffamily} & $k^{on}g_0^t$ & $k^{off}$ \\ [6pt]
  \textbf{Nonlinear\sffamily} & $k^{on}g_0^t\left(s_0-s\right)$ & $k^{off}s$ \\ [6pt]
\hline \\
\end{tabular*}
\vspace{12pt}

Then we define $r^{tot} = r^{on} + r^{off}$, $p^{off} = r^{off}/r^{tot}$, and $p^{on} = r^{on}/r^{tot}$.  The algorithm has the following steps:

\vspace{12pt}
\noindent \textbf{Repeat until $t>T_{max}$:}
\begin{enumerate}
 \item Choose a uniform random number $r_1 \in [0,1]$
 \item If $p^{off}<r_1$ then $\Delta = -1$, else $\Delta = 1$
 \item $s^t \to s^t + \Delta$, $g_1^t \to g_1^t -\Delta/dy$          
 \item Choose another uniform random number $r_2 \in [0,1]$
 \item $\dt = -log(r_2)/r^{tot}$
 \item $t = t + \dt$
 \end{enumerate}

\noindent The master equation for the surface describes the evolution of a discrete number of particles, while the PDE, Eq. \eqref{eqn:With_Source}, describes the evolution of concentration.  So, when a single particle is exchanged between the surface phase and the bulk phase, we exchange an amount $\Delta/\dy$ of concentration.

\subsection{Hybrid: KMC surface model with finite-difference bulk diffusion}
The hybrid model consists of alternating the KMC steps described above, and the diffusion steps described in Sec. \ref{sec:finite_difference}.  As described in Sec. \ref{sec:math_model}, we treat the KMC surface process as a source for the diffusion step.  The surface process (stochastic or deterministic) requires the concentration at the surface.  Thus we can say that the KMC step feeds concentration to the FEFD model, and the FEFD step feeds a concentration to the KMC model \footnote{If we had used the Robin boundary condition formulation, Sec. \ref{sec:math_model}, the KMC would supply a \emph{flux} for the finite difference portion.}.  In addition, the KMC step produces its own random time step.  Being random, this time step can be much smaller or much larger than the time step required for stability in the diffusion step.  Therefore, we imposed a lower and higher threshold for the diffusion time step $\dt_{low} = \alpha_{low}\tfrac{\dy^2}{D}$ and $\dt_{high} = \alpha_{high}\tfrac{\dy^2}{D}$; $\alpha_{low}<\alpha_{high}<1/4$.  For each KMC step, we perform anywhere from none to many diffusion time steps.  In this following algorithm, $\Delta^* t$ is the cumulative time since the last diffusion step.

\begin{enumerate}
 \item $\Delta^*t = 0$
 \item Do KMC step $\rightarrow \dt$
 \item $\Delta^*t = \Delta^*t + \dt$
 \item \textbf{if} $\Delta^*t < \dt_{low}$
  \begin{enumerate}
   \item Do not do a diffusion step
  \end{enumerate}
 \item \textbf{elseif} $\Delta^*t > \dt_{high}$
   \begin{enumerate}
    \item $M = [\Delta^*t/\dt_{high}]$
    \item $\delta t = \dt/M$
    \item do $M$ diffusion steps with $\delta t$
    \item $\Delta^* t = 0$
   \end{enumerate}
 \item \textbf{else} 
   \begin{enumerate}
    \item do 1 diffusion step with $\Delta^*t$
    \item $\Delta^* t = 0$ 
   \end{enumerate}
 \item go to 2.
\end{enumerate}

\section{Results \label{sec:results}}

\subsection{Validation -- linear model \label{sec:results_validation}}
\begin{figure*}[ht]
\centering

\subfigure[$k^{on} = 0.001$, $k^{off} = 10^-1$]{
   \includegraphics[height=2in]{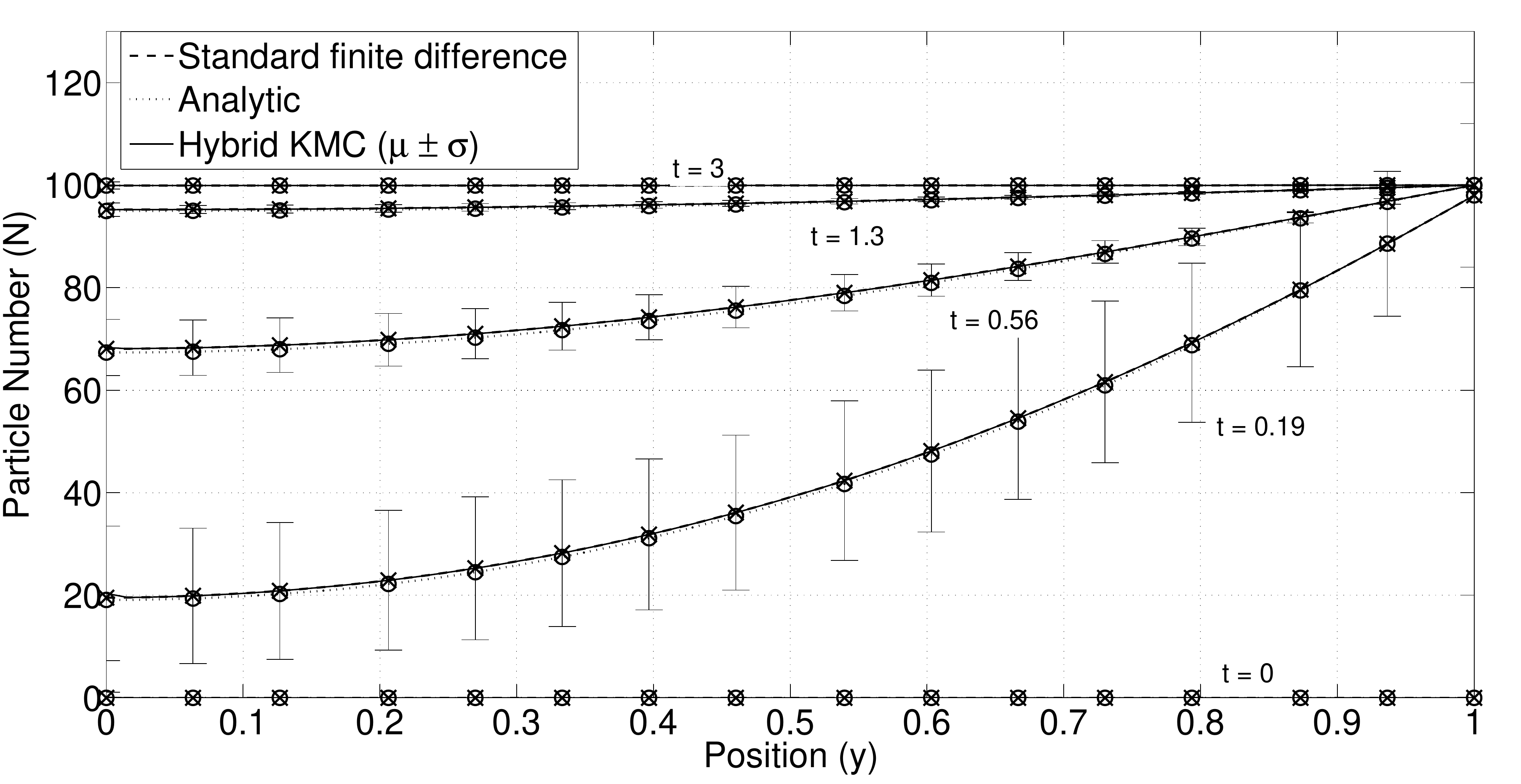}
   \label{fig:linear1}
 }

 \subfigure[$k^{on} = 0.01$, $k^{off} = 1$]{
   \includegraphics[height=2in]{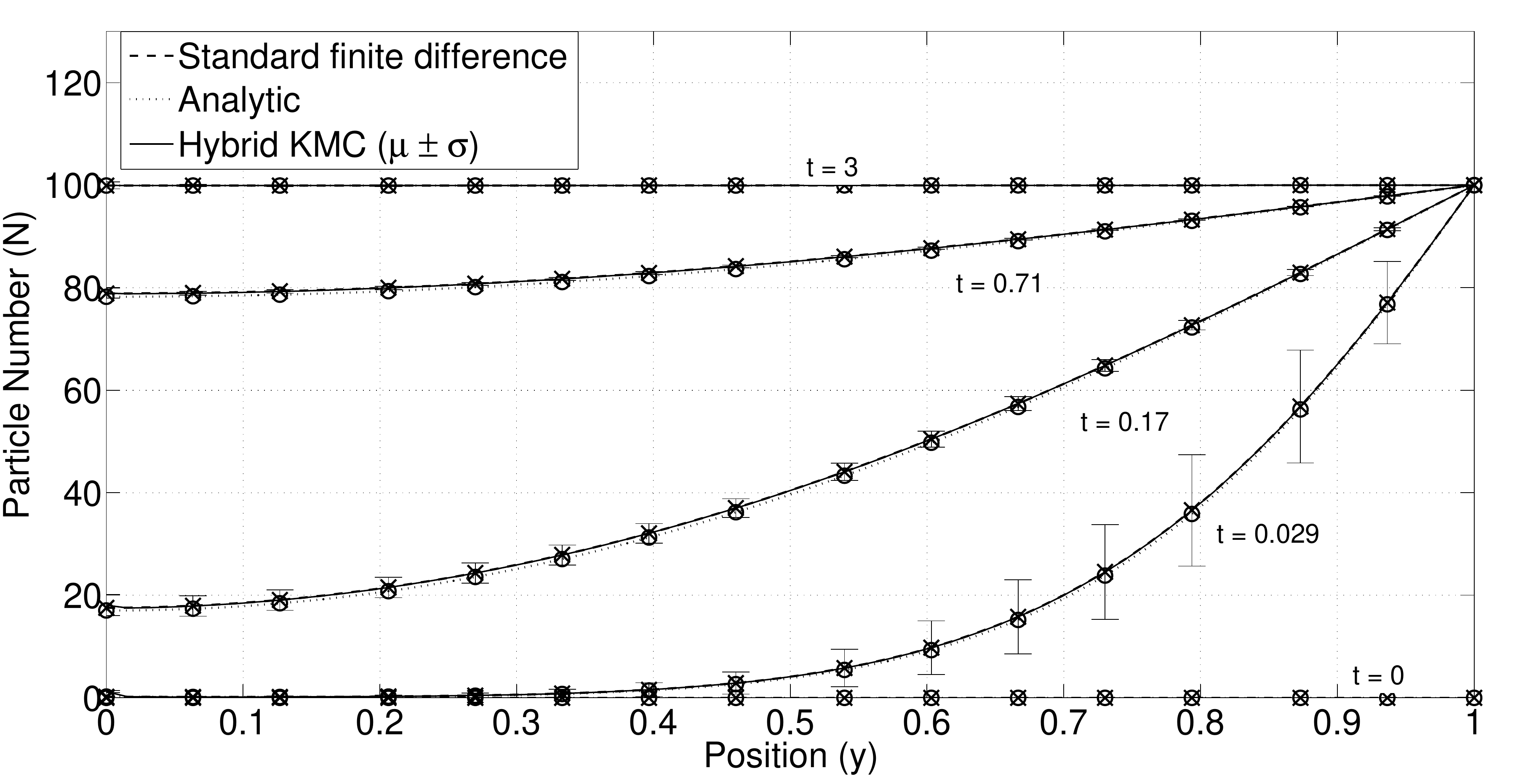}
   \label{fig:linear2}
 }

 \subfigure[$k^{on} = 0.1$, $k^{off} = 10$]{
   \includegraphics[height=2in]{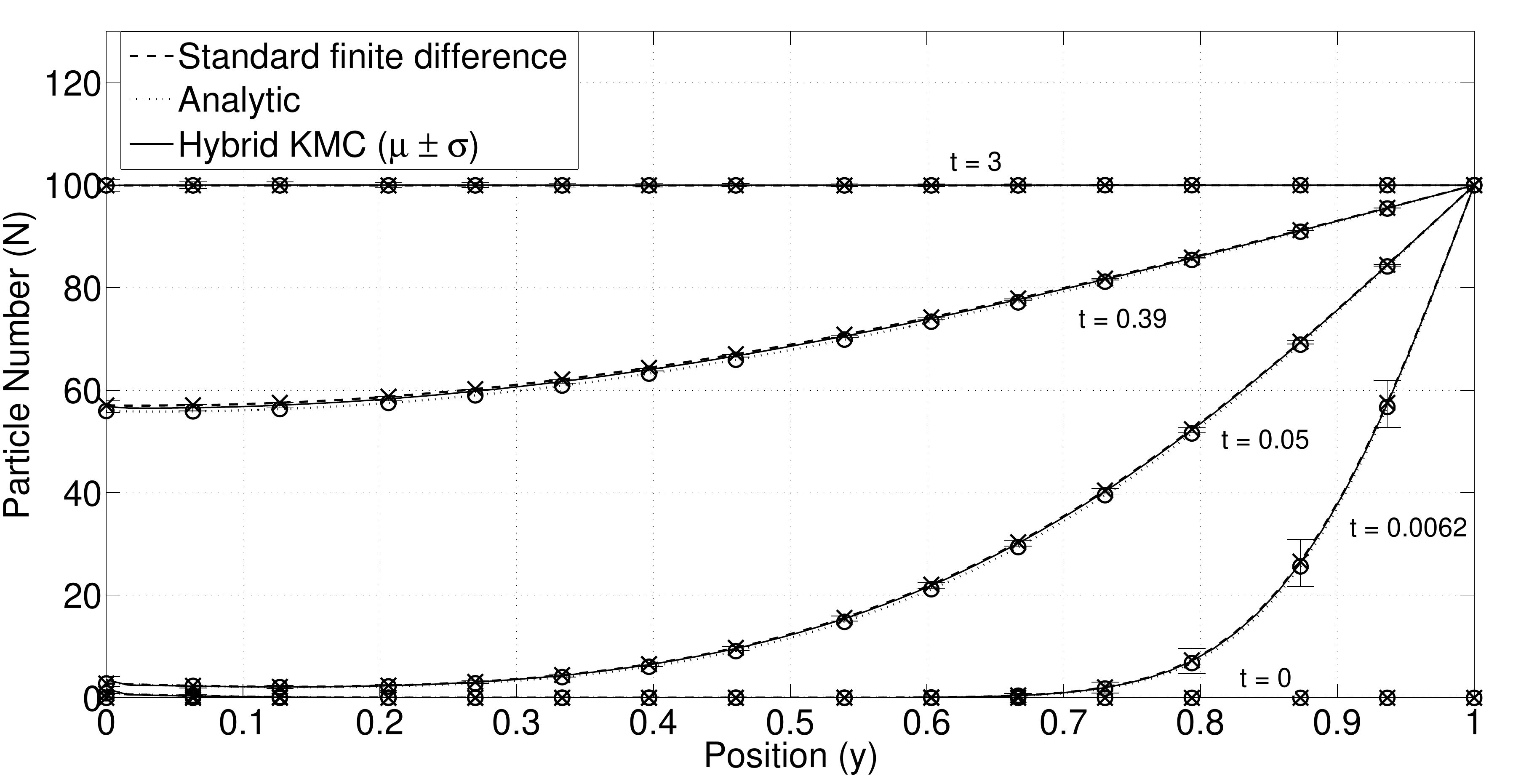}
   \label{fig:linear3}
 }
\caption{\label{fig:linear_time_dep} Time dependent solutions of the linear model with the initial bulk phase concentration zero everywhere.  KMC-continuum: solid line with ensemble fluctuations.  Deterministic continuum: dashed.  Analytic: dotted.  Concentration field is in units of particle number, $N(y) = g(y)\dy$.  (Ensemble size: $N=100$, $nY=10^2$, $Y=1$, $D=1$)}
\end{figure*}

Fig. \ref{fig:linear_time_dep} shows time dependent solutions of the linear model, Eqs. \eqref{eqn:With_Source} and \eqref{eqn:linear_source_rx}, with the initial bulk phase concentration zero everywhere.   Notice that we present the concentration field  in units of particles, $N(x,t) = g(y,t)\dy$ because our simulations are in a realm of small particle number.  This enables simpler and more intuitive comparison of the concentration at the surface, $g(0,t)$ with the number of particles on the surface $s(t)$. Fig. \ref{fig:linear_time_dep} shows a clear agreement between the analytical solution, the deterministic FEFD solution, and the hybrid KMC/FEFD solution (ensemble mean, N=100).  This demonstrate the accuracy of our numerical method.

\subsection{Nonlinear System  \label{sec:results_nonlinear}}
For the nonlinear system, we started from initial conditions in which the surface coverage was 0, and the bulk phase was initialized to the steady-state concentration $g_Y$.  The reason we did not look at evolution from $g(x,0) = 0$ in the bulk phase (such as we did for the linear model) is that in the nonlinear model, the surface process rate $R(s(t),g(0,t))$ is 0 until the gas reaches the surface (the linear rate law does not depend on $s(t)$).  In a KMC model, a rate of 0 corresponds to an infinite waiting time.  To accommodate this, we would need to suspend KMC steps until gas had diffused to the surface.  Rather than running simulations with diffusion only until $g$ diffused to the surface at $y=0$, it seemed reasonable to choose an initial condition that allowed the process to ensue immediately with a non-zero rate.  A corresponding physical model would be that the solution was allowed to equilibrate to a fixed concentration, and then the catalytic surface was suddenly immersed in the solution.

Pursuing this idea, we chose parameters with some physical relevance.  We set the surface capacity to $s_0 = 10^4$ particles.  We might think of this as a square surface $100$ molecules on a side, or approximately $1 nm$.  If we choose the physical dimensions of $\dy$ to also be $1 nm$, then the length of a domain with $Y=100$ is $0.1 \mu m$.  With these dimensions, a concentration of $1/\dy^3$ corresponds to $1 \, \text{mol}/\text{l}$.  Dimensional analysis gives $k=D/(L^2 g_Y)$, and the diffusion constant of a particle of molecular size is roughly $\approx 10^{-13}\,m^2/s$, giving values of $k \approx 10^{-6} \text{ to } 10^{-2} \, \text{l} \cdot \text{mol}^{-1} \cdot s^{-1}$ for the range of $D$ and $Y$ we simulated.  Reaction rates a few orders of magnitude on either side of this are very reasonable on a physical basis.  Therefore, while we chose parameter values that enabled us to run stable simulations in a reasonable amount of time, our parameters are also within reason for real systems.

The time dependent solution to these models, when looked at over the entire spatial domain, become excessively busy.  The concentration initially drops and the empty surface fills, and then increases back to the equilibrium value.  This caused the spatial concentration fields at different times to obscure each other.  In our nonlinear, catalysis model, the ensemble average concentration must equilibrate to the deterministic value.  However, as we discussed, the number of particles on the surface does not.  Therefore, for clarity and simplicity, we show only the values of $s(t)$ and $g(0,t)$.

%Figs \ref{fig:Control_Rx} to \ref{fig:Control_N_High_D} show results from simulations in which we fixed all parameters but one. The graphs give the ensemble average and ensemble fluctuations for ensembles of 100 independent realizations.  Fluctuations are the driving feature in our study.  Diffusion-induced fluctuations scale as $\sqrt{Dc}$ where $c$ is a characteristic concentration.  Reaction-induced fluctuations scale as $\sqrt{R}$ where $R$ is a characteristic reaction rate.  For instance, in our case this would be $R \sim \sqrt{k^{on}g(s-s_0) + k^{off}s}$ (this uses the additive property of the variances of Gaussian distributions).  The ratios $\sqrt{Dc}/Dc$ and $\sqrt{R}/R$ show that fluctuations become more important when the diffusion constant and reaction rates are small.  We can judge differences between different parameter sets in different ways, such as the rate at which the concentration equilibrates, and the ratio between the KMC-hybrid and mean field results during the transient zone.
%
%These considerations are consistent with $Q_1$ and $Q_2$ being larger, i.e. the systems being more reaction limited when the reaction rates are larger, and the gas concentration is smaller.

We have shown that when the stochastic and deterministic versions of the nonlinear model give different results, this is due to the presence of fluctuations.  In our models, fluctuations result from the stochastic nature of the chemical reactions.  However, these stochastic reactions are also coupled to diffusion.  When fluctuations enter the system, via the reactions, faster than they can diffuse away, then the system is in a reaction-limited regime.  It is in this regime that we expect the system to respond most strongly to fluctuations.

\begin{figure*}[ht]
\centering

\subfigure[$k^{on} = 0.0001$, $Q_1 = 10^{-1}$, $Q_2 = 10^3$]{\includegraphics[width=0.8\textwidth]{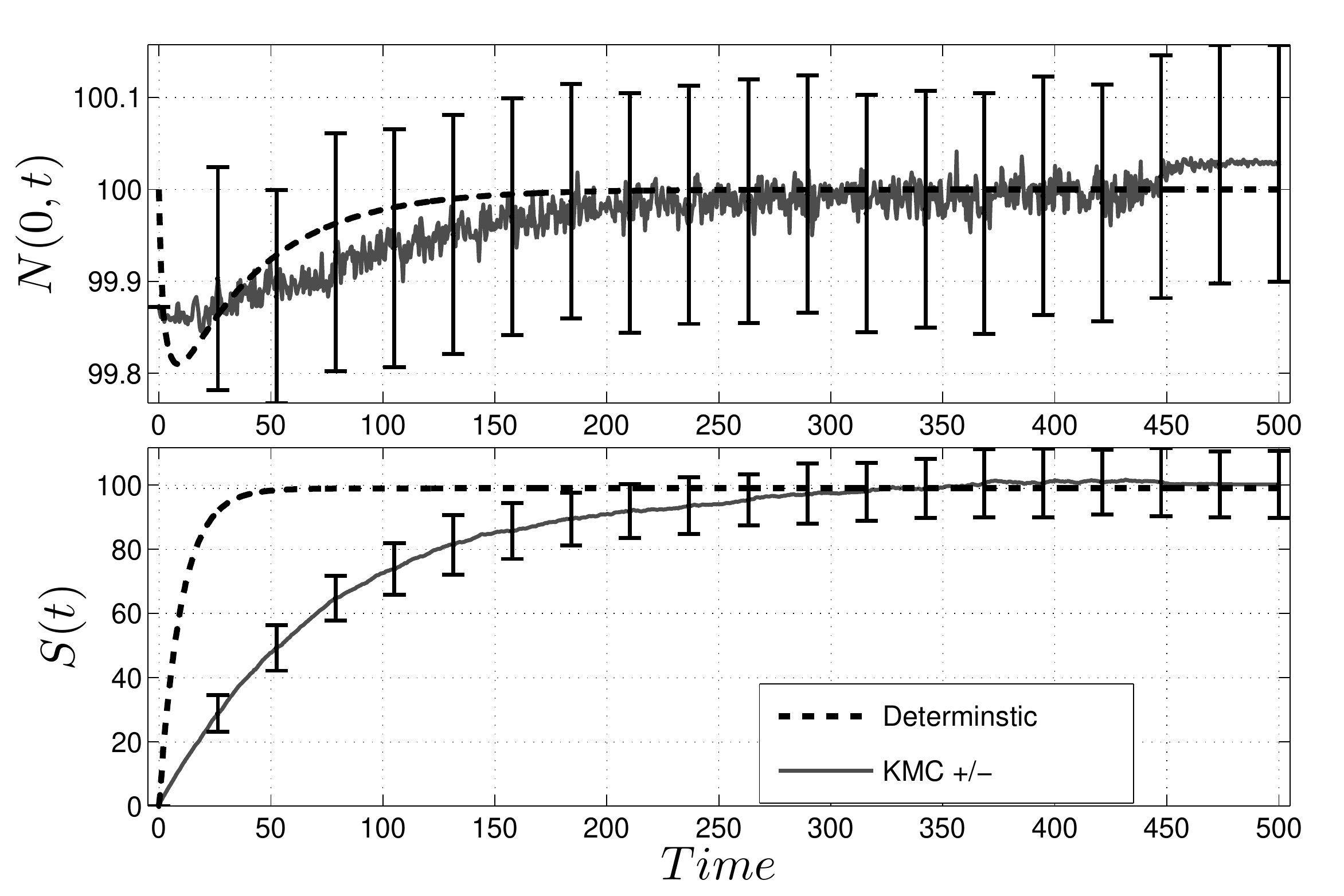}
 \label{fig:low_rx}
}

 \subfigure[$k^{on}=0.01$, $Q_1 = 10$, $Q_2 = 10^3$]{\includegraphics[width=0.8\textwidth]{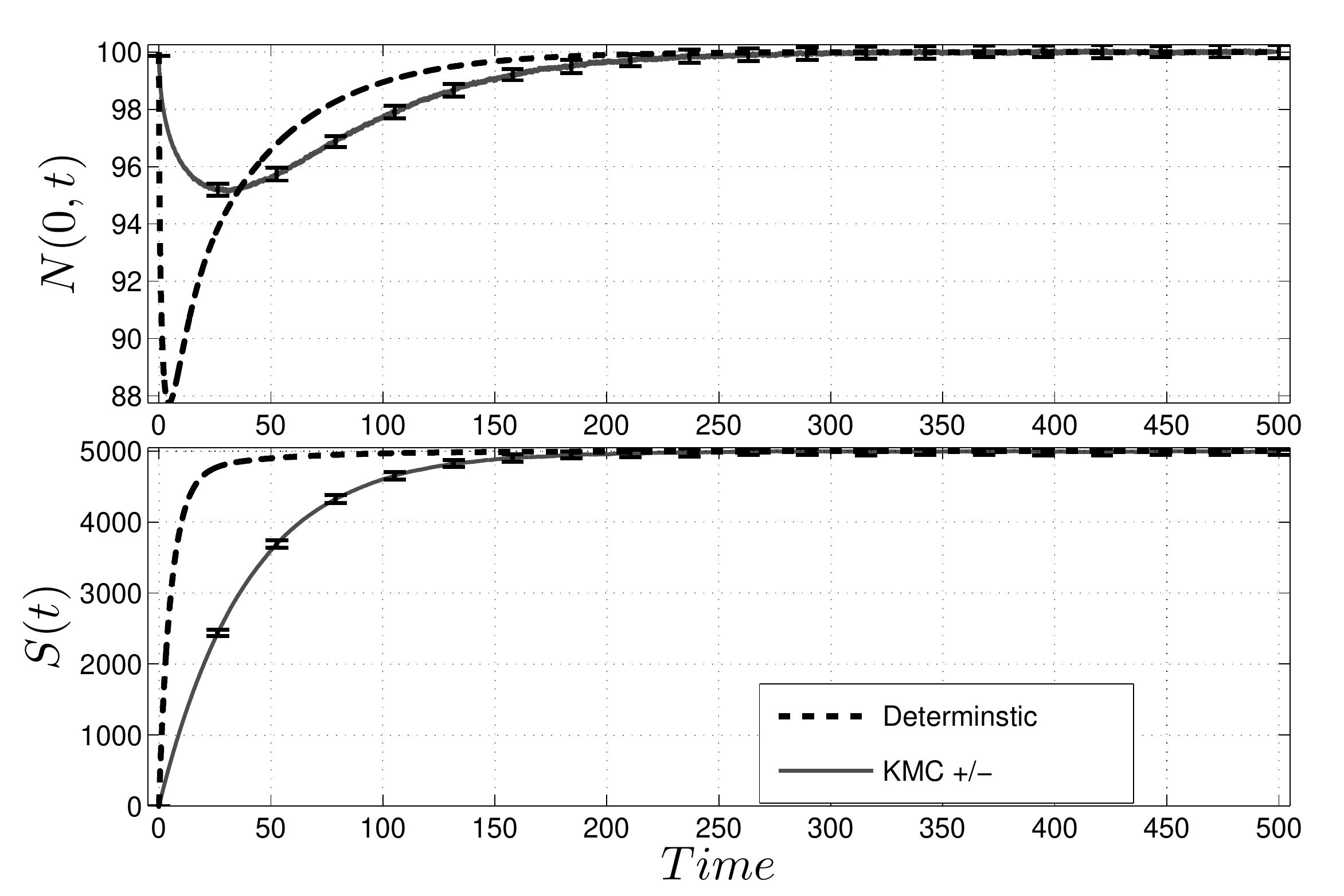}
 \label{fig:high_rx}
 }

\caption{\label{fig:Control_Rx} Nonlinear model with all parameters fixed except the on rate $k^{on}$.  (Ensemble size: $N=100$, $nY=10^2$, $S_0=10^4$, $k^{off}=10^-1$, $Y=10^3$, $D=10^4$)}
\end{figure*}

\begin{figure*}[ht]
\centering

 \subfigure[$D = 1000$, $Q_1 = 10^{-2}$, $Q_2 = 10^2$]{\includegraphics[width=0.8\textwidth]{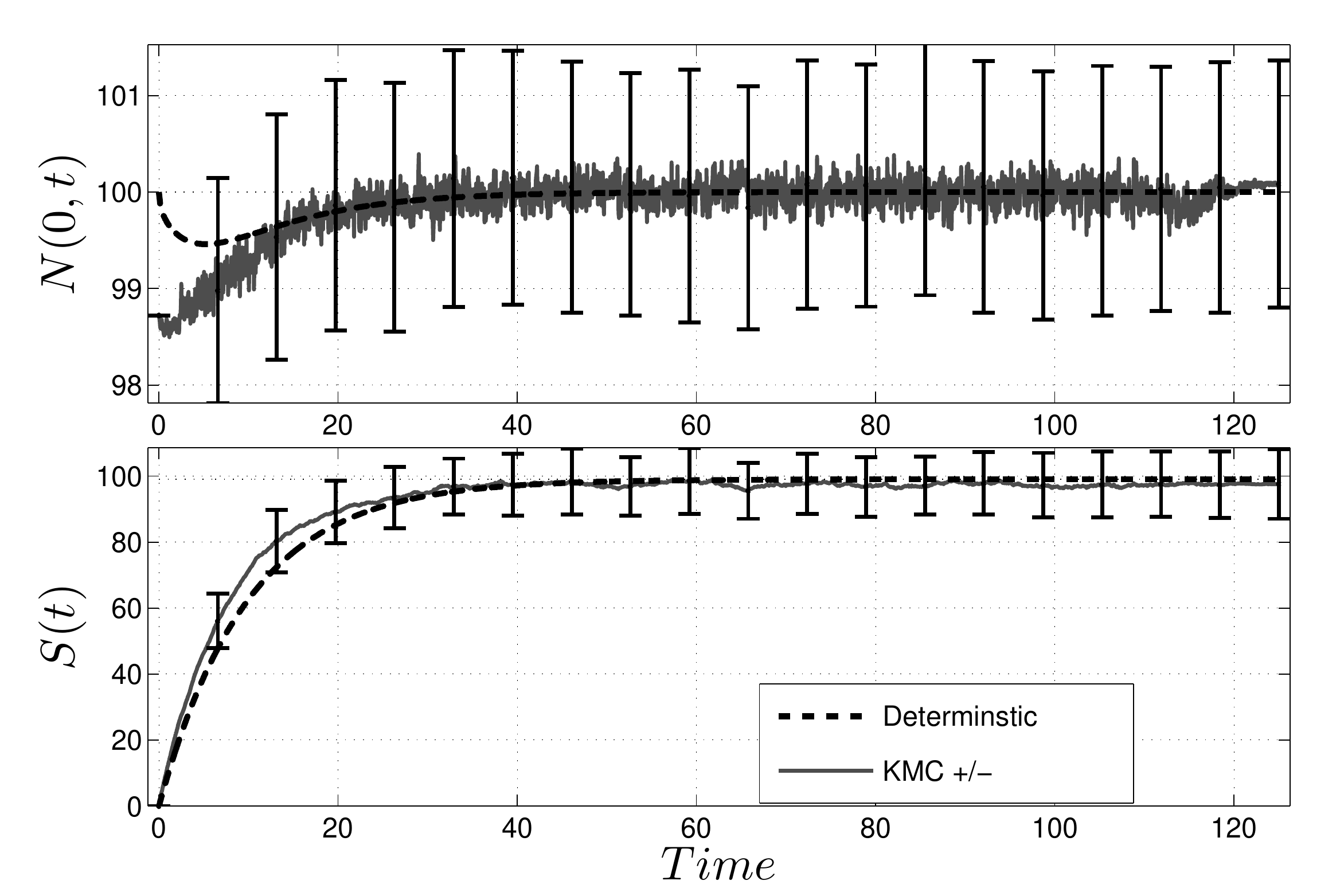}
 \label{fig:high_D}
 }

\subfigure[$D = 0.01$, $Q_1 = 10^3$, $Q_2 = 10^7$]{\includegraphics[width=0.8\textwidth]{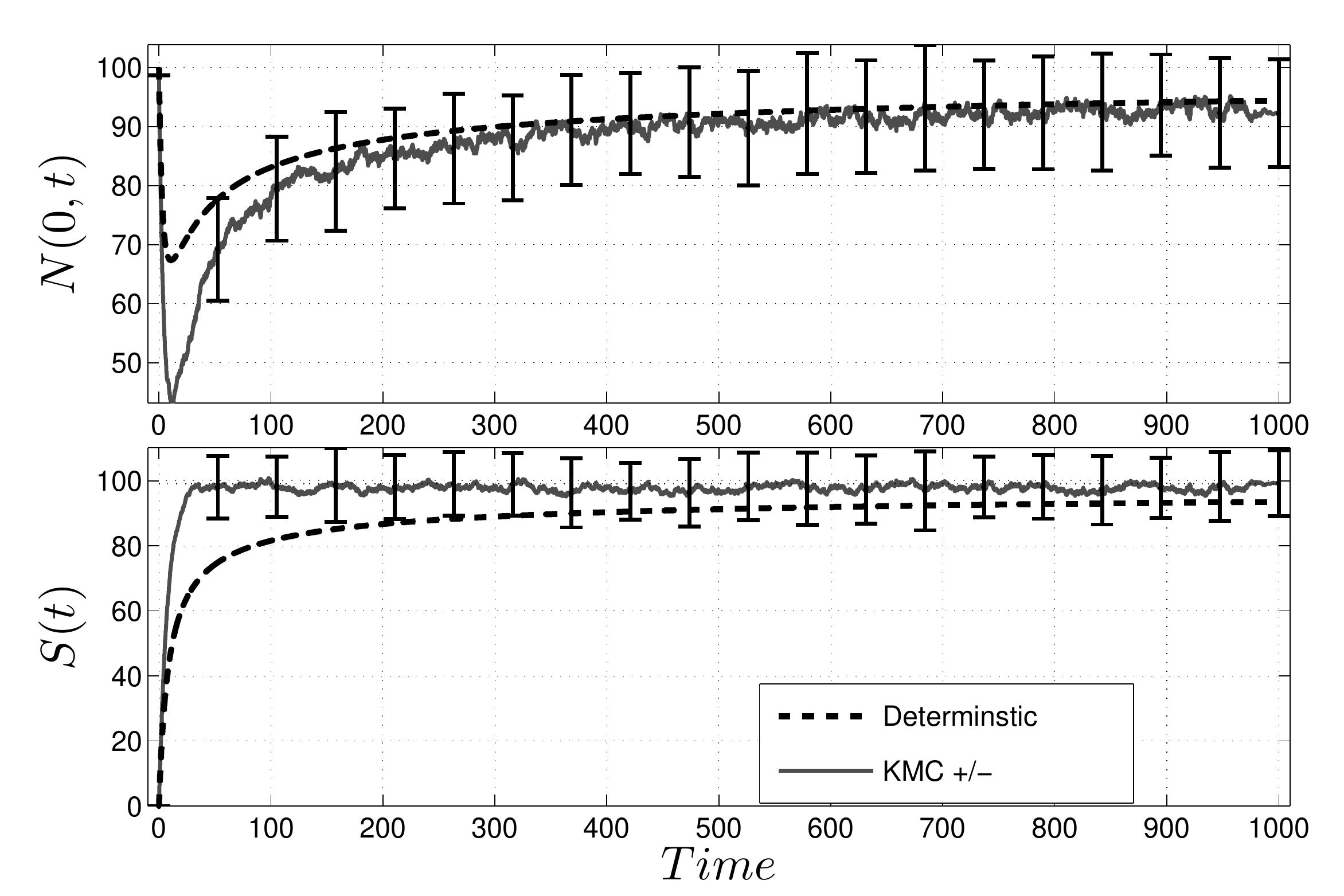}
 \label{fig:low_d}
 }

\caption{\label{fig:Control_D} Nonlinear model with all parameters fixed except the diffusion constant $D$.  Time and memory limitations prevented us from running the simulation for Fig. \ref{fig:low_d} until equilibrium.  (Ensemble size: $N=100$, $nY=10^2$, $S_0=10^4$, $k^{on}=10^{-5}$, $k^{off}=10^{-1}$, $Y=10^2$)}
\end{figure*}

\begin{figure*}[ht]
\centering

 \subfigure[$n_Y=10$, $Q_1 = 1$, $Q_2 = 10^6$]{\includegraphics[width=0.8\textwidth]{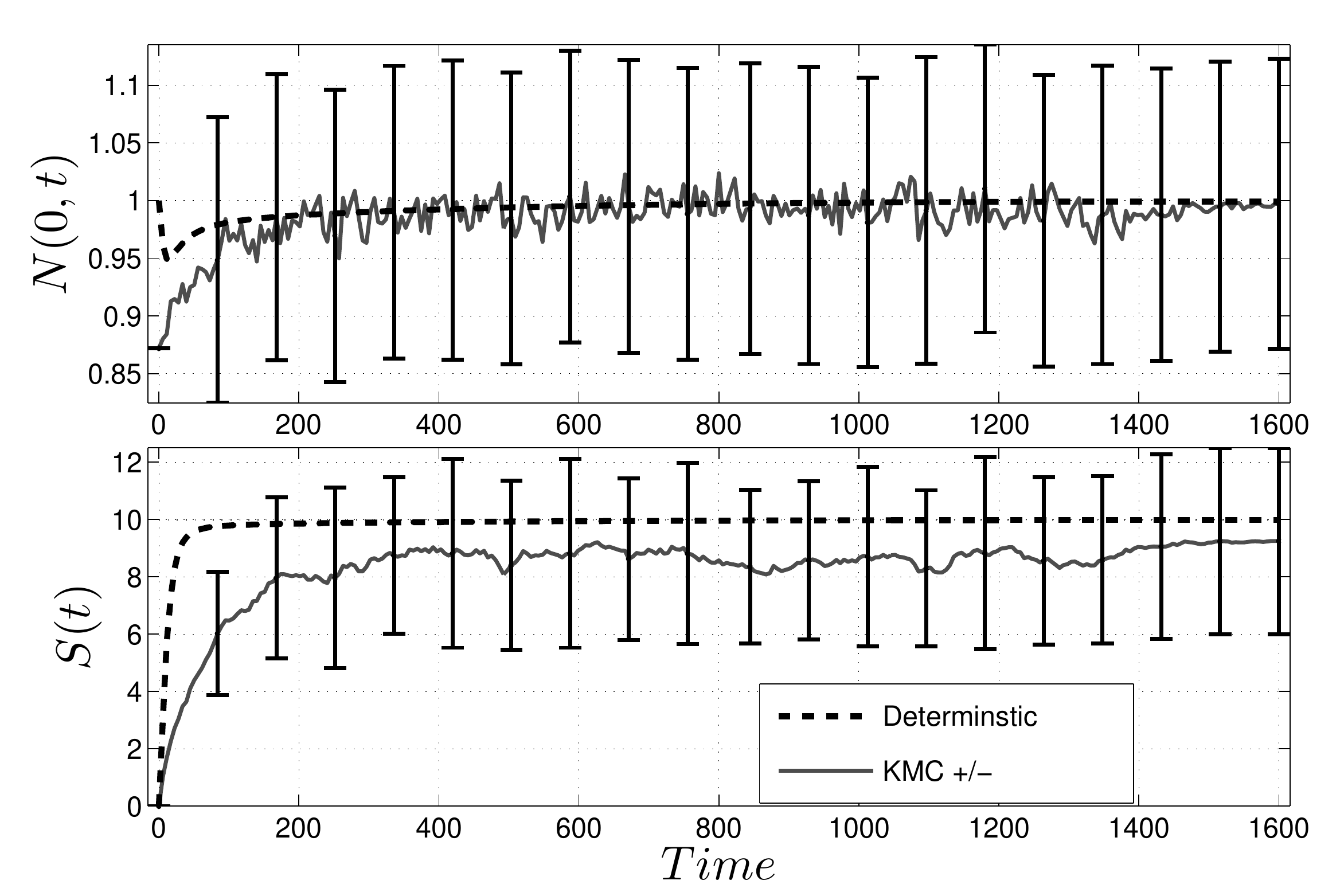}
 \label{fig:low_gY}
 }

\subfigure[$n_Y = 10000$, $Q_1 = 1$, $Q_2 = 10^4$]{\includegraphics[width=0.8\textwidth]{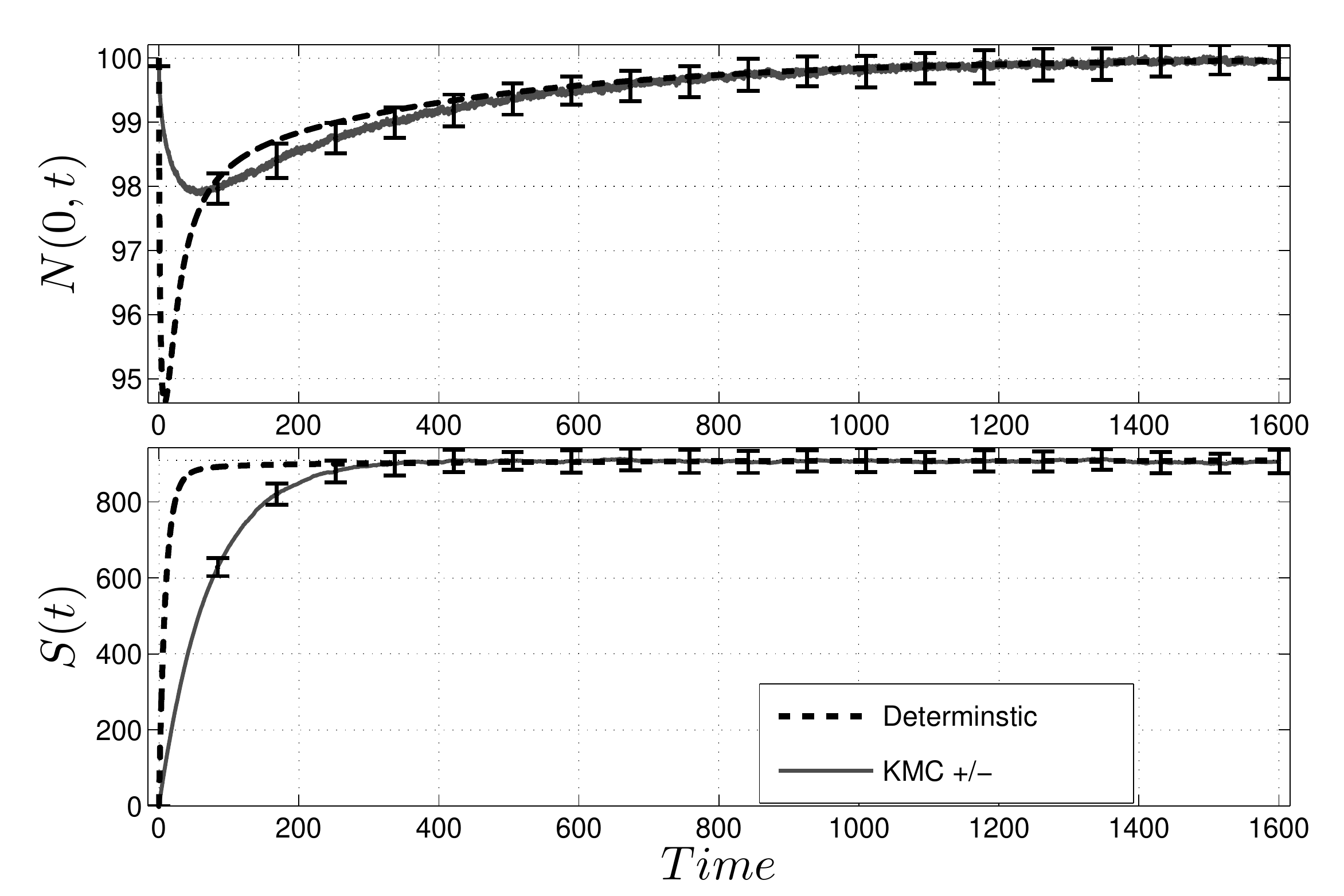}
 \label{fig:high_gY}
 }

\caption{\label{fig:Control_N_High_D} Nonlinear model with all parameters fixed except the fixed concentration $n_Y$.  (Ensemble size: $N=100$, $S_0=10^4$, $k^{on}=10^{-4}$, $k^{off}=10^{-1}$, $D=10^3$, $Y=10^3$)}
\end{figure*}

Figs \ref{fig:Control_Rx} to \ref{fig:Control_N_High_D} show results from simulations in which we fixed all parameters but one. The graphs give the ensemble average and ensemble fluctuations for ensembles of 100 independent realizations. 

To assess the effects of nonlinearity, we look at two qualitative measures in the graphs: (1) the initial dip in gas concentration during the transient, and (2) the steady-state surface concentration.  The initial dip in gas concentration occurs because, in accordance with our proposed physical situation (see section \ref{sec:results_nonlinear}), the catalytic surface is initially empty.  The gas concentration initially dips as material adsorbs onto the surface, but then recovers to the steady-state value.  The size of this dip, relative to the steady-state concentration is one difference we see as we make the system more reaction-limited.  We have discussed in section \ref{sec:stoch_splitting} that due to the nonlinearity, the steady-state surface concentration in the master-equation model and the mean-field model.  The size of this effect also becomes more pronounced as we make the system more reaction-limited.

For instance, in Fig. \ref{fig:Control_D}, Fig. \ref{fig:high_D} both $Q_1$ and $Q_2$ are much larger than in Fig. \ref{fig:low_d}.  We also see a much larger transient dip.  In Fig. \ref{fig:Control_N_High_D} $Q_2$ is much larger in Fig. \ref{fig:low_gY} where we see a much greater relative difference in the steady-state surface concentration.

\section{Conclusion \label{sec:conclusion}}
We have successfully validated a hybrid numerical model that couples KMC for a surface process to finite difference for bulk diffusion.  We have demonstrated that a KMC surface model coupled to a finite difference bulk diffusion model can exhibit behavior not predicted by a deterministic counterpart when the surface process rate law is nonlinear.  We first validated a numerical method by comparing results of a linear model to an analytic solution.  Then, we showed, through stochastic analysis of moment equations, that the observed difference is not a product of the numerical method, but a fundamental property of the coupled nonlinear system.  We used this analysis to successfully explain our observed results.

This study highlights some important concepts:  The evolution of a system modeled with deterministic mean-field surface model is different than with a KMC surface model, this may have two causes.  This difference may arise from the mean-field surface model being unable to capture the complexity of the surface processes.  But it may also arise when there is nonlinear coupling between the surface process and the gas phase -- simply because of the nonlinear coupling.  Even when a deterministic mean-field model for a surface process is available, if the rate equations are nonlinear, then one should still use a KMC driven surface model to obtain the most realistic results.  Such a case might be when there are only weak interactions between solid phase molecules -- a physical situation similar to the models we studied.  

\section{Acknowledgements \label{sec:acknowledgements}}
This work was financially supported by the Laboratory Directed Research and Development (LDRD) project at Pacific Northwest National Laboratory (PNNL) and Applied Mathematics program of the US DOE Office of Advanced Scientific Computing Research. The Pacific Northwest National Laboratory is operated by Battelle for the U.S. Department of Energy under Contract DE-AC05-76RL01830.

\bibliographystyle{plain}
\bibliography{hybrid2}

\end{document}